\documentclass[
showkeys,
reprint,
superscriptaddress,
nofootinbib,
nobibnotes,
longbibliography,
aps,
prb,
floatfix,
nofootinbib
]{revtex4-2}

\usepackage{soul}
\usepackage{amsmath}
\usepackage{amssymb}
\usepackage{amsfonts}
\usepackage{tablefootnote}
\usepackage{graphicx}
\usepackage{color}
\usepackage[dvipsnames]{xcolor}
\usepackage{bbm}
\usepackage[breaklinks=true,colorlinks=true,linkcolor=blue,urlcolor=blue,citecolor=blue]{hyperref}
\usepackage{array}
\usepackage{booktabs}
\usepackage{mathptmx}
\usepackage{placeins}

\DeclareMathAlphabet{\mathcal}{OMS}{cmsy}{m}{n}
\newcolumntype{P}[1]{>{\centering\arraybackslash}p{#1}}
\newcolumntype{M}[1]{>{\centering\arraybackslash}m{#1}}

\bibstyle{Apsrev4-2}

\begin{document}

\title{Annealing-induced grain coarsening and voltage kinks in superconducting NbRe films}

\author{Zahra Makhdoumi Kakhaki}
\email{z.makhdoumi-kakhaki@tu-braunschweig.de}
    \affiliation{Cryogenic Quantum Electronics, Institute for Electrical Measurement Science and Fundamental Electrical Engineering (EMG) and Laboratory for Emerging Nanometrology (LENA), Technische Universit\"at Braunschweig, 38106, Braunschweig, Germany}

\author{Anton O. Pokusinskyi}
    \affiliation{Cryogenic Quantum Electronics, Institute for Electrical Measurement Science and Fundamental Electrical Engineering (EMG) and Laboratory for Emerging Nanometrology (LENA), Technische Universit\"at Braunschweig, 38106, Braunschweig, Germany}

\author{Francesco Avitabile}
    \affiliation{Dipartimento di Fisica “E. R. Caianiello”, Università degli Studi di Salerno, I-84084 Fisciano, Salerno, Italy}
    
\author{Abhishek Kumar}
     \affiliation{Dipartimento di Fisica “E. R. Caianiello”, Università degli Studi di Salerno, I-84084 Fisciano, Salerno, Italy}
     \affiliation{CNR-SPIN, c/o Università degli Studi di Salerno, I-84084 Fisciano, Salerno, Italy}

\author{Francesco Colangelo}
       \affiliation{Dipartimento di Fisica “E. R. Caianiello”, Università degli Studi di Salerno, I-84084 Fisciano, Salerno, Italy}
       \affiliation{CNR-SPIN, c/o Università degli Studi di Salerno, I-84084 Fisciano, Salerno, Italy}
\author{Carla Cirillo}

        \affiliation{CNR-SPIN, c/o Università degli Studi di Salerno, I-84084 Fisciano, Salerno, Italy}
\author{Carmine Attanasio}
       \affiliation{Dipartimento di Fisica “E. R. Caianiello”, Università degli Studi di Salerno, I-84084 Fisciano, Salerno, Italy}
       \affiliation{CNR-SPIN, c/o Università degli Studi di Salerno, I-84084 Fisciano, Salerno, Italy}
         \affiliation{Centro NANOMATES, c/o Università degli Studi di Salerno, I-84084 Fisciano, Salerno, Italy
}

\author{Oleksandr V. Dobrovolskiy}
    \affiliation{Cryogenic Quantum Electronics, Institute for Electrical Measurement Science and Fundamental Electrical Engineering (EMG) and Laboratory for Emerging Nanometrology (LENA), Technische Universit\"at Braunschweig, 38106, Braunschweig, Germany}
    \affiliation{FLUXONICS---The European Foundry for Superconducting Electronics e.V., 38116 Braunschweig, Germany}

\begin{abstract}
NbRe, a non-centrosymmetric superconductor with a transition temperature $T_\mathrm{c}$ of up to about 9\,K, attracts interest for its strong antisymmetric spin-orbit coupling and suitability for single-photon detection. While bulk and thin-film polycrystalline NbRe are well studied, how superconductivity and vortex dynamics evolve with increasing grain size in thin films is largely unknown. Here, we investigate as-grown and annealed 20\,nm-thick NbRe films, where annealing increases the average crystallite size from approximately $2$\,nm to $8$\,nm, and probe their resistance states via current-voltage ($I$-$V$) measurements over a broad temperature and magnetic field range. In contrast to as-grown films, where the low-resistive state breaks down due to flux-flow instability, annealed films exhibit multiple voltage kinks in the $I$-$V$ curves. We attribute these kinks to the nucleation and growth of normal domains, as further suggested by time-dependent Ginzburg-Landau simulations. Overall, the annealed films form superconducting networks with vortex-channeling paths along the grain boundaries, while localized heating and voltage kinks may offer potential for discrete-resistance switching and sensing.
\end{abstract}
\maketitle

\section{Introduction}
Disordered Nb$_{0.18}$Re$_{0.82}$ (NbRe) thin films have attracted increasing attention for superconducting single-photon detectors~\cite{cirillo2020superconducting,ejrnaes2022single,cirillo2024single}, gate-tunable devices~\cite{koch2024gate}, and superinductors~\cite{battisti2024demonstration}. Bulk non-centrosymmetric NbRe crystallizes in the cubic $I\overline{4}3m$ structure and is characterized by strong antisymmetric spin-orbit coupling~\cite{Karki2011} and a superconducting (SC) transition temperature $T_\mathrm{c}$ of about 9\,K~\cite{Chen2013}. NbRe single crystals have been shown to exhibit multiband superconductivity~\cite{Cirillo2015}. Disordered NbRe thin films, despite exhibiting a single SC gap~\cite{Cirillo2016}, show spin-triplet correlations when coupled to ferromagnetic layers in spin-valve structures~\cite{colangelo2025unveiling}. Controlling the crystallite size in such films thus provides a route to tuning and understanding their SC properties.

Sputtered NbRe films are typically polycrystalline with grain sizes of $\sim2$\,nm and lie in the SC dirty limit~\cite{cirillo2022polycrystalline,tinkham2004introduction}. Their transport properties are governed by microstructural features such as crystallite size, crystallographic orientation, and disorder. Thermal annealing provides a means to tune the crystallite size~\cite{makhdoumi2024effect}, but systematic studies of annealing-induced changes in key SC parameters, including the upper critical field $B_\mathrm{c2}$, electron diffusion coefficient $D$, and coherence length $\xi$, remain scarce. Likewise, while recent studies on NbRe thin films have addressed vortex dynamics and flux-flow instabilities (FFIs)~\cite{caputo2017nbre,de2026fast,kakhaki2024characterization,Cirillo2016}, the role of crystallite size and microstructural disorder in the current-driven resistive transition remains largely unexplored.

In uniform materials with $D \simeq 0.5$\,cm$^2$/s and/or fast electron energy relaxation~\cite{Emb17nac,Dob20nac,budinska2022rising}, magnetic flux quanta (Abrikosov vortices) can reach high critical velocities of $v^\ast \gtrsim 10$\,km/s before FFI onset~\cite{larkin1975nonlinear}. In contrast, spatial inhomogeneities broaden the distribution of vortex velocities~\cite{Sil12njp}, reduce $v^\ast$~\cite{budinska2022rising}, and favor local rather than global FFI nucleation~\cite{bezuglyj2019local,Vod19sst}. With efficient heat removal, fast-moving vortices form vortex rivers, i.e., phase-slip lines with an oscillating SC order parameter~\cite{Sil10prl,Siv03prl}. If heat removal is insufficient, or if the transport current density, $J^\ast$, exceeds $J_\mathrm{eq}$---the maximum current density at which a nonisothermal normal-superconducting (N/S) boundary may vanish~\cite{bezuglyj2019local}---vortex rivers may turn into N domains which will expand, driving the film into the normal state. Film microstructure therefore plays a key role in vortex ordering, dissipation, and the mechanisms governing the resistive transition. Annealing-induced grain coarsening is furthermore expected to produce oxidation at grain boundaries, locally suppressing the SC order parameter and creating branched networks that facilitate vortex channeling and inhibit global FFI nucleation.

In this work, we compare the normal-state and SC properties of as-grown and annealed NbRe films and study vortex dynamics via current-voltage ($I$-$V$) measurements. Unlike single FFI jumps observed for as-grown films, annealed films display multiple voltage kinks in their $I$-$V$ curves.  On the one hand, annealed films exhibit higher resistivity, which is favorable for achieving the large kinetic inductance required in superinductors~\cite{battisti2024demonstration}. On the other hand, annealing-induced grain coarsening and oxidation at grain boundaries promote the onset of a resistive state at lower vortex velocities $v^\ast$, thereby limiting their suitability for fast fluxonic devices. Time-dependent Ginzburg-Landau (TDGL) simulations reproduce the main experimental features and indicate that the resistive transition in as-grown films is governed by FFI, whereas the multi-step $I$-$V$ curves of annealed films arise from the nucleation and growth of N domains along the grain boundaries. Our results highlight how microstructural tuning via annealing affects vortex dynamics, dissipation, and FFIs in NbRe thin films, while localized heating and voltage kinks may offer potential for discrete-resistance switching and sensing.

\section{Experiment}
\subsection{Samples}
\begin{figure}[t]
  \centering
  \includegraphics[width=8.6cm]{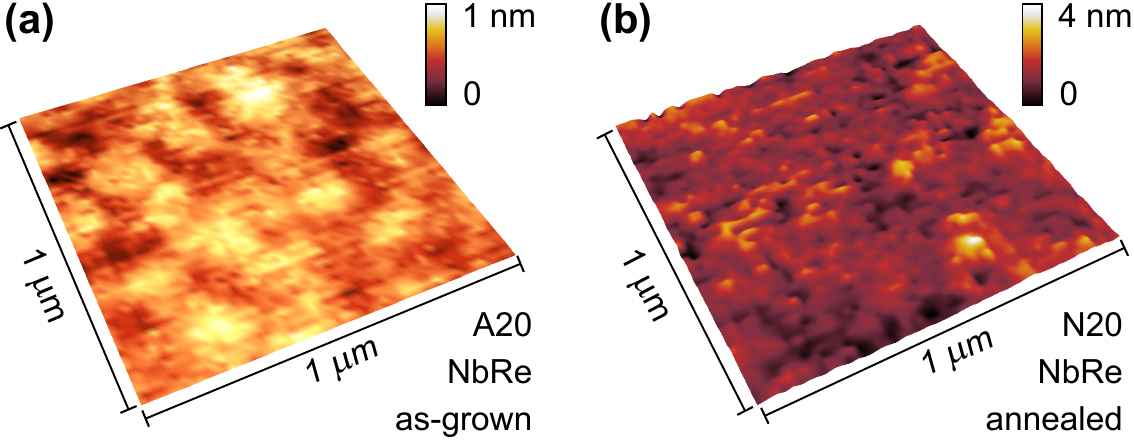}
  \caption{\label{AFM} AFM images of the surfaces of the as-grown (a) and annealed (b) NbRe films.}
  \label{fig:AFM}
\end{figure}

Our studies were carried out on $20$\,nm-thick NbRe films deposited by DC magnetron sputtering from a Nb$_{0.18}$Re$_{0.82}$ target onto Si/SiO$_2$ substrates at room temperature. The base pressure prior to deposition was in the low $10^{-8}$\,Torr range and sputtering was performed in an Ar atmosphere at $3$\,mTorr. The as-grown films were annealed at $600^\circ$C for $30$\,min in a tube furnace, followed by an additional annealing step at $300^\circ$C for $30$\,min. Previous X-ray diffraction analysis showed that annealing increases the grain size from $\sim 2$\,nm to $\sim 8$\,nm~\cite{makhdoumi2024effect}. Inspection of the film surface by atomic force microscopy (AFM) revealed a flat surface for as-grown films, with an RMS roughness of $\sim 0.7$\,nm over a $1\times1\,\mu\mathrm{m}^2$ area. In contrast,  annealed films exhibit a grainy morphology and increased roughness of $\sim 3.5$\,nm, see Fig.~\ref{fig:AFM}. 

For electrical transport measurements, one as-grown film (sample A20) and one annealed film (sample N20) were investigated. The temperature dependence of the resistance $R(T)$ was measured on unpatterned films under different perpendicular magnetic fields in order to determine the upper critical field $B_\mathrm{c2}(T)$. For the $I$-$V$ measurements, the films were patterned into four-probe geometries. The strip width was $W = 10\,\mu$m, while the lengths were $L = 100\,\mu$m for the as-grown sample A20 and $L = 80\,\mu$m for the annealed sample N20. All measurements were performed with the magnetic field applied perpendicular to the film plane.

$I$-$V$ characteristics were recorded in the current-driven regime at fixed temperatures and different magnetic fields. The vortex instability velocity $v^\ast$ was determined from the last voltage point $V^\ast$ at the onset of the resistive jump using the standard relation, $v^\ast = V^\ast/(B L)$, where $B$ is the magnetic flux density and $L$ is the distance between the voltage leads.

\subsection{Resistance temperature dependence}

Figure~\ref{Hc2}(a,b) shows the temperature dependence of the electrical resistance, $R(T)$, for the as-grown NbRe film A20 and annealed film N20 in magnetic fields ranging from 0 to 5\,T. For sample A20, the normal-state resistivity at 10\,K is $\rho_{10\mathrm{K}} = 143\,\mu\Omega\cdot\mathrm{cm}$, and the residual resistance ratio, defined as $\mathrm{RRR} = R_{300\mathrm{K}}/R_{10\mathrm{K}} \approx 0.95$, indicates a strongly disordered metallic state, as typically observed in sputtered NbRe thin films~\cite{Cirillo2016,makhdoumi2024effect}. Using the 50\% resistance criterion, the SC transition temperature is $T_\mathrm{c}^{\mathrm{A20}} = 6.60\,\mathrm{K}$, slightly higher than that of the annealed sample N20, for which $T_\mathrm{c}^{\mathrm{N20}} = 6.28\,\mathrm{K}$. The main geometrical and electrical parameters of the samples are summarized in Table~\ref{tab:combined_parameters}.

\begin{table}[t!]
\centering
\caption{Geometrical, electrical, superconducting, and vortex dynamics parameters for as-grown and annealed NbRe films.}
\label{tab:combined_parameters}
\begin{tabular}{lcc}
\hline
\textbf{Parameter} & \textbf{A20} & \textbf{N20} \\
\hline

\multicolumn{3}{c}{\textit{Geometrical and electrical parameters}} \\
Thickness, $d$ (nm) & 20 & 20 \\
Width, $W$ ($\mu$m) & 10 & 10 \\
Length, $L$ ($\mu$m) & 100 & 80 \\
Resistivity at 10\,K, ($\mu\Omega\cdot$cm) & 143 & 330 \\
RRR & 0.95 & 1.13 \\

\multicolumn{3}{c}{\textit{Superconducting parameters}} \\
$T_\mathrm{c}$ (K) & 6.60 & 6.28 \\
$D$ ($\mathrm{cm}^2/\mathrm{s}$) & 0.55 & 0.57 \\
$\xi(0)$ (nm) & 6.0 & 6.3 \\
$\left.\dfrac{dB_{\mathrm{c2}}}{dT}\right|_{T_\mathrm{c}}$ (T/K) & 2.0 & 1.9 \\
$B_\mathrm{c2}(0)$ (T) & 10 & 13 \\
$B_\mathrm{c1}(0)$ (mT) & 3.3 & 1.7 \\
$\lambda(0)$ (nm) & 492 & 765 \\
$\Lambda(0)$ ($\mu$m) & 23 & 60 \\

\multicolumn{3}{c}{\textit{Vortex dynamics parameters}} \\
$v^{\ast}$ at low field (km/s) & $1.4$ & $0.7$ (kink~0) \\
$v^{\ast}$ at saturation field (km/s) & $0.6$ & $0.1$ (kink~0) \\
\hline
\end{tabular}
\end{table}

After annealing, the normal-state resistivity increases significantly to $\rho_{10\mathrm{K}} = 330\,\mu\Omega\cdot\mathrm{cm}$, while the residual resistance ratio increases to $\mathrm{RRR} \approx 1.13$. The increase in RRR suggests a modest reduction in temperature-independent scattering, consistent with some improvement in crystalline order, likely due to grain coarsening and partial defect healing during annealing. At the same time, the substantial increase in resistivity indicates the presence of additional scattering mechanisms, which we attribute to interdiffusion at the film--substrate interface and oxidation at grain boundaries, in agreement with previous reports~\cite{makhdoumi2024effect}. These competing effects result in a transport behavior that remains overall strongly disordered, despite the slight increase in RRR. The reduction of $T_\mathrm{c}$ after annealing is consistent with these additional sources of disorder and interface degradation.

The magnetic-field dependence of the SC transition width, defined as $\Delta T_\mathrm{c} = T_{90\%R} - T_{10\%R}$, is presented in Fig.~\ref{Hc2}(c). For both samples, the transition width increases with increasing magnetic field, reflecting enhanced vortex motion and dissipation near the SC transition. Notably, the annealed film exhibits a smaller $\Delta T_\mathrm{c}$ and a weaker dependence on magnetic field compared to the as-grown film. This behavior suggests improved homogeneity and possibly enhanced SC properties at the grain level, consistent with the partial structural ordering induced by annealing. 

\subsection{Upper critical field}
\begin{figure}[t]
  \centering
  \includegraphics[width=8.6cm]{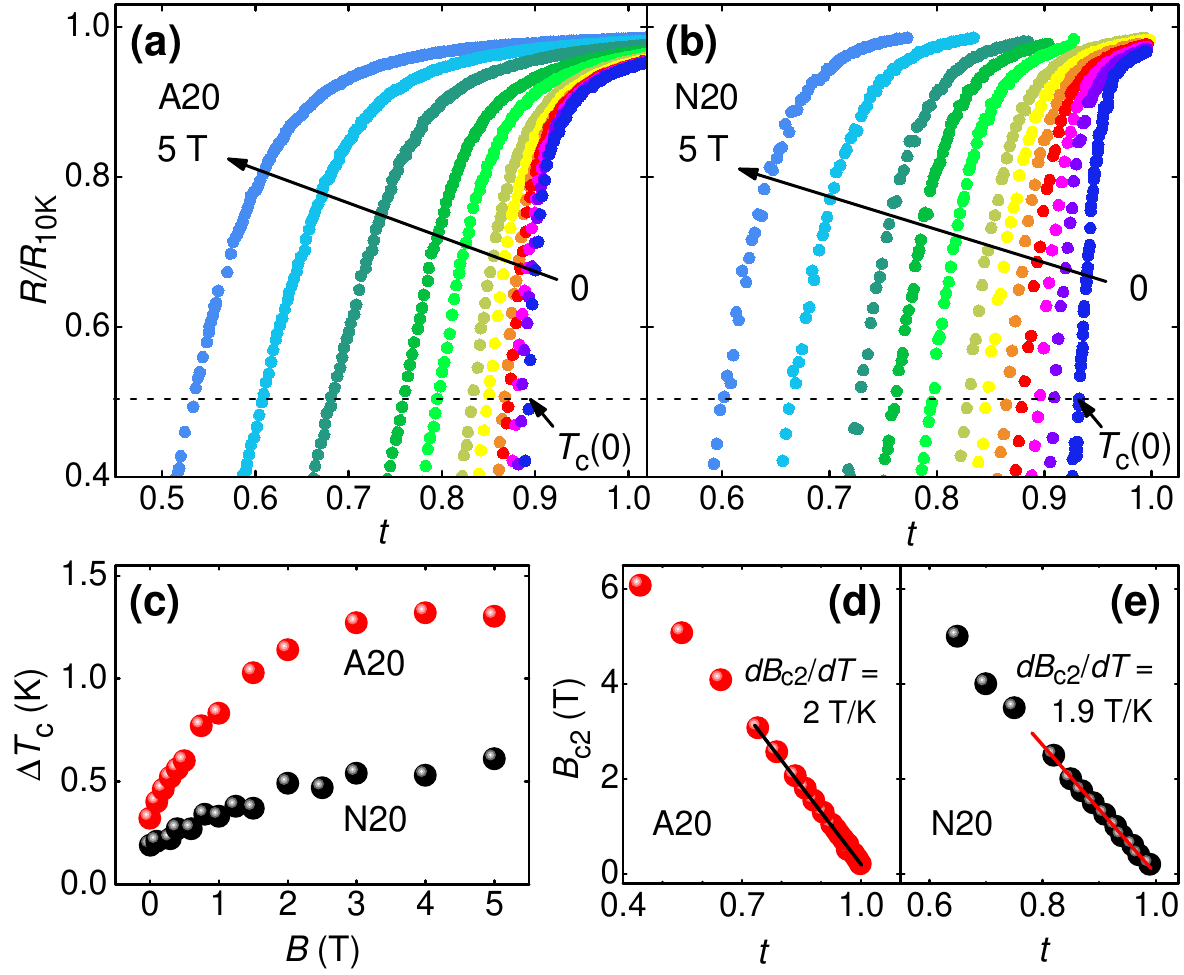}
  \caption{
  Temperature dependence of the normalized resistance for the as-grown (a) and annealed (b) NbRe films at magnetic fields ranging from $0$ to $5$\,T.
  Magnetic-field dependence of the superconducting transition width $\Delta T_\mathrm{c}$ (c) and temperature dependence of the upper critical field $B_\mathrm{c2}(T)$ (d,\,e) for the same samples.
  Symbols: experiment; lines in (d,\,e): linear fits near $T_\mathrm{c}$.}
  \label{Hc2}
\end{figure}

The temperature dependence of the upper critical field, $B_\mathrm{c2}(T)$, extracted from the $R(T)$ curves using the 50\% resistance criterion, is shown in Fig.~\ref{Hc2}(d,\,e) for samples A20 and N20, respectively. For both films, $B_\mathrm{c2}(T)$ decreases approximately linearly with temperature in the vicinity of $T_\mathrm{c}$. The slopes $\left. dB_\mathrm{c2}/dT \right|_{T_\mathrm{c}}$ are found to be very similar for the two samples. Within a single-band dirty-limit framework~\cite{tinkham2004introduction}, this observation suggests comparable electron diffusion coefficients $D$ and SC coherence lengths $\xi$. This result is notable in light of the structural modifications induced by annealing. While grain coarsening is expected to improve crystallinity and reduce disorder within individual grains, the overall transport properties indicate increased resistivity after annealing. This apparent discrepancy can be reconciled by considering the coexistence of competing effects: improved intra-grain order and enhanced SC properties at the microscopic level, alongside increased inter-grain scattering due to oxidation and interdiffusion at grain boundaries. As a consequence, the effective SC length scales, which depend on the global electronic transport, remain nearly unchanged.

Using the dirty-limit relations $\xi(0) = \sqrt{\hbar D / (1.76 k_\mathrm{B} T_\mathrm{c})}$
and $D = 4k_\mathrm{B}(\pi e\left| dB_{\mathrm{c}2}/dT \right|_{T_\mathrm{c}})^{-1}$, 
we estimate the zero-temperature coherence lengths to be $\xi_{\mathrm{A20}}(0) \approx 6.0~\mathrm{nm}$ for the as-grown film and $\xi_{\mathrm{N20}}(0) \approx 6.3~\mathrm{nm}$ for the annealed film. These similar values are consistent with the comparable slopes of $B_\mathrm{c2}(T)$ near $T_\mathrm{c}$. The London penetration depth at zero temperature is estimated using the dirty-limit expression $
\lambda(0) = 1.05 \times 10^{-3} \sqrt{\rho_{10\,\mathrm{K}} / T_{\mathrm{c}}}$, 
yielding $\lambda_{\mathrm{A20}}(0) \approx 492~\mathrm{nm}$ and $\lambda_{\mathrm{N20}}(0) \approx 765~\mathrm{nm}$. The corresponding Ginzburg--Landau parameters, $\kappa = \lambda / \xi$, are $\kappa \approx 82$ for A20 and $\kappa \approx 121$ for N20, indicating that both films lie deep in the type-II  regime. The Pearl lengths, $\Lambda(0) = 2 \lambda^2(0)/d$,
are $\Lambda(0) \approx 23\,\mu\mathrm{m}$ and $\Lambda(0) \approx 60\,\mu\mathrm{m}$, respectively. These values confirm that the films are in the thin-film limit ($d \ll \lambda$) and satisfy $\xi \ll w \lesssim \Lambda$.

The lower critical field at zero temperature is estimated using
$B_\mathrm{c1}(0) = \frac{\Phi_0}{4\pi \lambda^2(0)} \left( \ln \kappa + 0.5 \right)$, where $\Phi_0$ is the magnetic flux quantum. This yields $B_\mathrm{c1}(0) \approx 3.3\,\mathrm{mT}$ for A20 and $B_\mathrm{c1}(0) \approx 1.7\,\mathrm{mT}$ for N20. The zero-temperature upper critical field $B_{c2}(0)$ is estimated using the interpolation formula~\cite{Karki2011}, $B_{\mathrm{c}2}(0)=B_{\mathrm{c}2}(T)(1+t^{2})/(1-t^{2})$, 
which reproduces the linear Ginzburg-Landau behavior near $T_\mathrm{c}$ and captures the expected low-temperature curvature. Fits to the experimental data (not shown) yield $B_\mathrm{c2}(0) \approx 10\,\mathrm{T}$ for A20 and $B_\mathrm{c2}(0) \approx 13~\mathrm{T}$ for N20. Despite the higher $B_\mathrm{c2}(0)$ of the annealed film, the extracted coherence lengths remain very similar. This indicates that the simple dirty-limit relationship between $B_\mathrm{c2}$ and $\xi$ does not fully capture the underlying physics in these films. The enhancement of $B_\mathrm{c2}(0)$ upon annealing may instead arise from a combination of factors, including changes in scattering mechanisms, possible multiband effects, or modifications of the electronic structure. A more detailed analysis of the annealing-induced evolution of $B_\mathrm{c2}(T)$ will be the subject of future work.

\subsection{Current-voltage curves}
\begin{figure}[t]
  \centering
  \includegraphics[width=8.6cm]{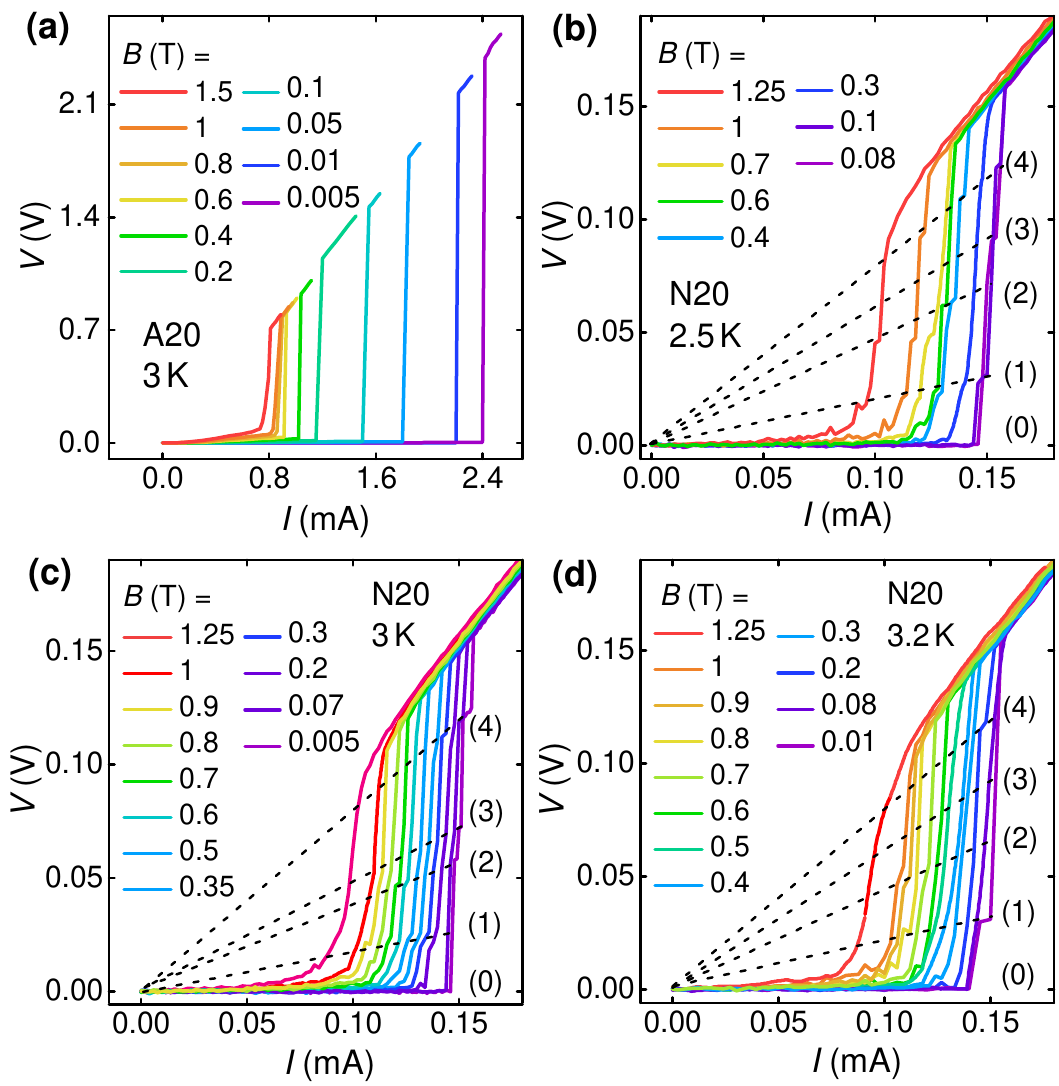}
  \caption{\label{I-V}
  $I$-$V$ curves for as-grown sample A20 at $3$\,K (a) and annealed sample N20 at $2.5$\,K (b), $3.0$\,K (c), and $3.2$\,K (d), for a range of applied magnetic fields. The dashed lines in panels (b)-(d) are drawn to highlight the voltage kinks and unequal spacing between them.}
\end{figure}
\begin{figure*}[t]
  \centering
  \includegraphics[width=0.92\textwidth]{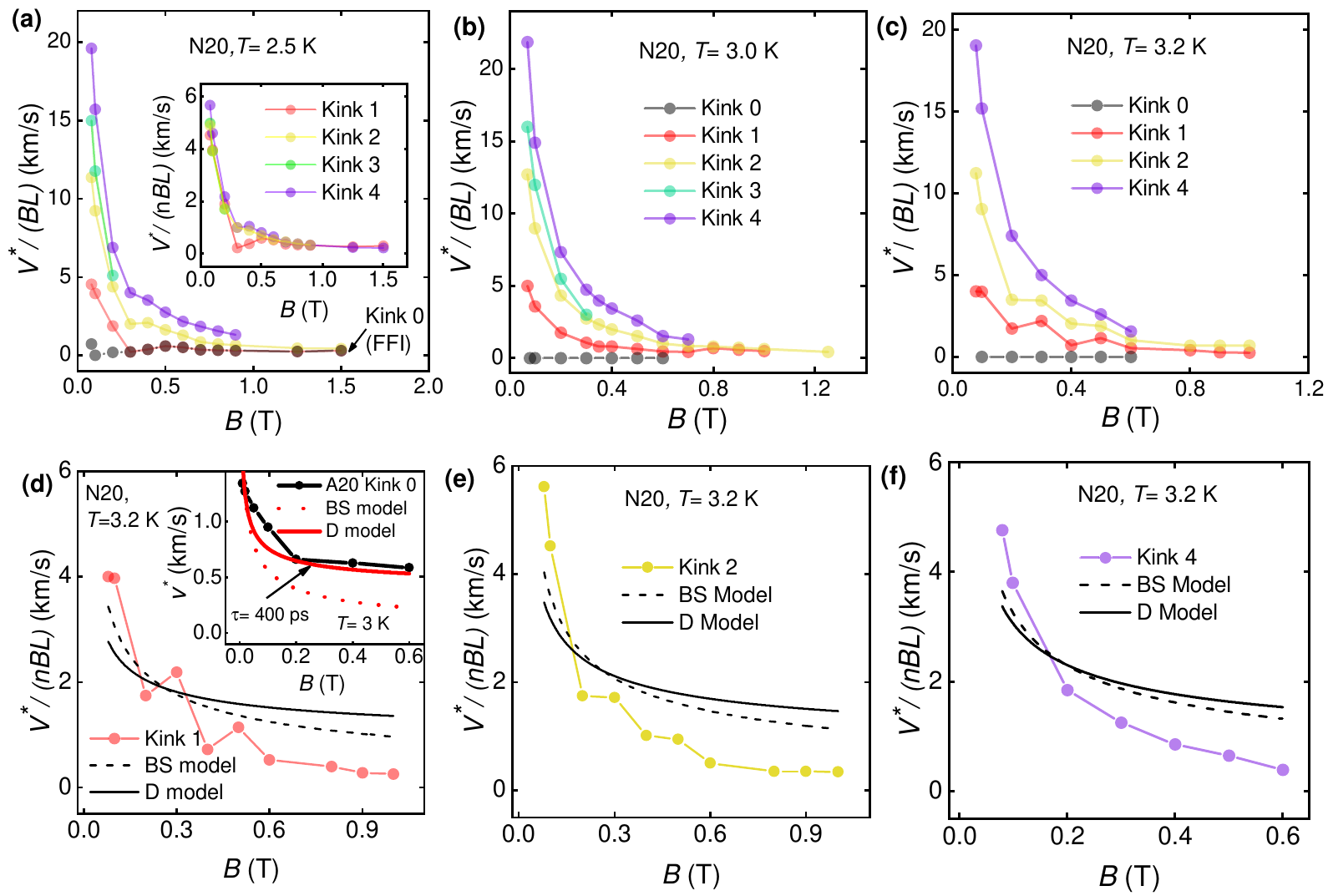}
  \caption{
  (a-c) Kink voltages $V^{\ast}/(BL)$ as a function of magnetic field for kink orders $n = 0$--$4$ at $T = 2.5$, $3.0$, and $3.2$\,K for the annealed film N20. Note that at $T = 3.2$\,K the kink with $n = 3$ is not observed.
  (d-f) Experimental data (symbols) and fits to the Doettinger (D, solid lines) and Bezuglyi-Shklovskij (BS, dashed lines) models at $3.2$\,K.
  The inset in (d) shows $v^\ast(B)$ and fits to Eqs.\,\eqref{eq:LO} and\,\eqref{eq:BS} for film A20 at $3$\,K.}
  \label{Velocity}
\end{figure*}

Figure~\ref{I-V}(a) shows the $I$-$V$ curves for a series of magnetic fields for sample A20 at $3$\,K while Figs.~\ref{I-V}(b-d) display the $I$-$V$ curves for sample N20 for three different temperatures. For magnetic fields up to approximately $1.5$\,T, sample A20 exhibits abrupt FFI jumps, whereas such single, well-defined jumps are absent for sample N20. Instead, annealed film N20 displays up to four distinct voltage kinks during the transition to the normal state. With increasing magnetic field, the higher-order kinks (kinks~3 and~4) progressively disappear, while signatures of kinks~1 and~2 persist at higher fields. The lines drawn from the origin to the kinks in Figs.~\ref{I-V}(b)-(d) highlight that the kinks are spaced unequally.

From the last current point preceding the first voltage jump and the subsequent kinks, we deduce the characteristic voltages $V^\ast$. For the first jump, we associate $V^\ast$ with the FFI voltage~\cite{larkin1975nonlinear,Dob24inb}. Further possible origins of $V^\ast$ associated with the voltage kinks are discussed in Sec.~\ref{sec:discussion}. Figures~\ref{Velocity}(a-c) show the ratios $V^{\ast}/(BL)$ for sample N20 as a function of the magnetic field at $T=2.5$, $3.0$, and $3.2$\,K, respectively. At all temperatures, $V^{\ast}/(BL)$ is largest at low fields and decreases monotonically with increasing magnetic field. The inset of Fig.~\ref{Velocity}(a) shows the kink voltage normalized by the kink order, $V^{\ast}/(nBL)$, at $2.5$\,K, where $n$ is the kink index, indicating that $V^\ast$ does not scale linearly with $n$. This behavior is also apparent in the $I$-$V$ curves of Fig.~\ref{I-V}(b)-(d), where the slopes of the dashed lines for the voltage kinks vary, and the differences between consecutive slopes are unequal.

Within FFI models, the ratio $V^{\ast}/(BL)$ corresponds to the vortex instability velocity $v^\ast$. In the Doettinger (D) model, $v^\ast$ decreases with increasing magnetic field according to
\begin{equation}
    v^{\ast}(B,\tau_\mathrm{E}) =
    [14\,\zeta(3)\left(1 - t\right)]^{1/4}
    \sqrt{\tfrac{D}{\pi\,\tau_\mathrm{E}}}
    \left[
    1 + \tfrac{1}{\sqrt{D\,\tau_\mathrm{E}}}
    \sqrt{\tfrac{2\Phi_0}{\sqrt{3}\,B}}
    \right],
\label{eq:LO}
\end{equation}
where $\tau_\mathrm{E}$ is the electron energy relaxation time and $t=T/T_\mathrm{c}$. The D model generalizes the Larkin–Ovchinnikov (LO) framework~\cite{larkin1975nonlinear} by accounting for the finite quasiparticle diffusion length relative to the intervortex spacing set by the applied magnetic field. However, neither the LO nor the D model includes the finite rate of heat removal mediated by the escape of nonequilibrium phonons into the substrate. This limitation is addressed in the self-heating model developed by Bezuglyi and Shklovskij (BS) \cite{bezuglyj1992effect}, which predicts
\begin{equation}
    v^\ast(B) = C_\mathrm{BS}/\sqrt{B}.
\label{eq:BS}
\end{equation}

Figures~\ref{Velocity}(d-f) show fits of $V^\ast(B)/(BL)$ at $T = 3.2$\,K to the above models. The deduced $v^{\ast}$ data for as-grown sample A20 range from $1.4$\,km/s at low fields to $0.7$\,km/s at $0.5$\,T, with the field dependence $v^{\ast}(B)$ satisfactorily described by the D model using an electron energy relaxation time $\tau_\mathrm{E} = 400$\,ps as a fitting parameter, see the inset of Fig.~\ref{Velocity}(d). These $v^{\ast}$ and $\tau_\mathrm{E}$ values are of the same order of magnitude as $v^{\ast} \simeq 0.3-1.2$\,km/s and $\tau_\mathrm{E} \simeq 700$\,ps reported for 15\,nm-thick MoSi strips~\cite{budinska2022rising} with $T_\mathrm{c} \approx 6.43$\,K and rough edges at $0.78T_\mathrm{c}$. However, since investigating the magnetic-field dependence of the critical current at very low fields---and thus assessing the edge quality of the studied constrictions---is beyond the scope of this work, we do not associate the deduced $\tau_\mathrm{E} = 400$\,ps with the intrinsic properties of the as-grown NbRe films.

By contrast, the voltage-kink data for the annealed sample N20 shown in Fig.~\ref{Velocity}(d--f) deviate significantly from both the D and BS models, indicating that the instability in this sample is governed by a different mechanism. In particular, the experimental $V^{\ast}/(nBL)$ values decrease much more rapidly with magnetic field than predicted by either model. The deduced $V^{\ast}/(nBL)$ reaches values of about $20$\,km/s at low fields; however, interpreting this quantity as a vortex velocity is not justified for spatially non-uniform superconducting films~\cite{bezuglyj2019local}. 

This observation reflects a more general point: fast vortex motion is not the only dynamic state accessible at high currents and is, in fact, unlikely in nonuniform, coarse-grained disordered superconductors, as discussed below. Accordingly, the notation $V^{\ast}/(nBL)$ in Fig.~\ref{Velocity}(d--f) is used deliberately to emphasize that, despite its units, the ratio $V^{\ast}/(BL)$ should not be interpreted as a direct measure of the vortex velocity, even when its magnitude falls within a seemingly reasonable range of several km/s.

Namely, due to the annealing-induced order-parameter suppression at grain boundaries, FFI is expected to nucleate not in the entire sample but only in some areas with faster vortex motion \cite{bezuglyj2019local}. This should result in local overheating and the formation of normal (N) domains. If $J^\ast > J_\mathrm{eq}$, where $J_\mathrm{eq}$ is the current density corresponding to the equilibrium of a nonisothermal N/S boundary \cite{Gur84spu,Bez84ltp}, the entire film will transition to the normal state via the growth of N domains. By contrast, if $J^\ast < J_\mathrm{eq}$, nonstationary N domains will nucleate and subsequently vanish within the film \cite{Bez84ltp}. Thus, for the dynamic state with moving vortices, $J^\ast$ should be smaller than 
\begin{equation}
    J_\mathrm{eq} =
    \left[
    \frac{2h\,(T_\mathrm{c} - T)}{R_{\square}\,d^{2}}
    \right]^{1/2},
    \label{eq:Jeq}
\end{equation}
where $h$ is the heat-removal coefficient, $R_{\square}$ the sheet resistance in the normal state, and $d$ the film thickness. For the studied NbRe films $h$ is unknown; yet its upper-bound estimate can be taken for film-substrate interfaces with efficient heat removal, such as epitaxial (110) Nb films on ($11\bar{2}0$) sapphire substrates. Substituting the parameters of sample N20, $\rho = 330\,\mu\Omega$\,cm, $d = 20$\,nm, $(T_\mathrm{c} - T) = 3.78$\,K, and $h_{\mathrm{Nb/Al_2O_3}} = 0.27 \times 10^{4}\,\mathrm{W\,m^{-2}\,K^{-1}}$ \cite{bezuglyj2019local} into Eq. \eqref{eq:Jeq} yields $J_\mathrm{eq} \simeq 56$\,kA/cm$^2$. This value is lower than the typical current densities at which voltage kinks are observed in our experiments, where $J^\ast(0\,\mathrm{T},\,2.5\,\mathrm{K}) \simeq 66$\,kA/cm$^2$. Consequently, assuming that the actual smaller heat-removal coefficient for the NbRe/SiO$_2$ interface further strengthens the inequality $J^\ast > J_\mathrm{eq}$, the voltage kinks observed in the annealed state must be triggered by the nucleation and growth of N domains. Accordingly, the values $V^\ast/(BL)$ reported for annealed film N20 in Fig.\,\ref{Velocity} should not be interpreted as vortex velocity. While some correspondence between the magnetic-field dependence of $V^\ast/(BL)$ and the $v^\ast(B) \sim B^{-1/2}$ scaling in the BS model may be attributed to the thermal nature of the underlying processes, the observed faster decrease of $V^\ast/(BL)$ suggests additional effects. In particular, this discrepancy likely arises from the spatially nonuniform nucleation of N domains and the specifics of heat conduction within the film and its removal to the substrate. 

For sample A20, $J_\mathrm{eq} \simeq 88\,\mathrm{kA/cm^2}$, which is much smaller than $J^\ast(0\,\mathrm{T},\,3\,\mathrm{K}) \simeq 1\,\mathrm{MA/cm^2}$. In this case, however, dissipation occurs uniformly across the sample, and the high dissipated power, $\sim \rho_\mathrm{f} J^{\ast 2}$, is governed primarily by $J^\ast$ rather than by the flux-flow resistivity $\rho_\mathrm{f}$. In contrast, film N20 exhibits a $\rho_{10\,\mathrm{K}}$ that is approximately twice as large and a measured $J^\ast$ that is an order of magnitude smaller than those in film A20. Owing to the significant contribution of grain boundaries to $\rho_{10\,\mathrm{K}}$, dissipation in sample N20 is highly nonuniform, and the measured $J^\ast$ is substantially reduced compared to the value expected for a SC state of the grains in the absence of grain boundaries.

\subsection{Critical current}
\begin{figure}[t!]
  \centering
  \includegraphics[width=8.6cm]{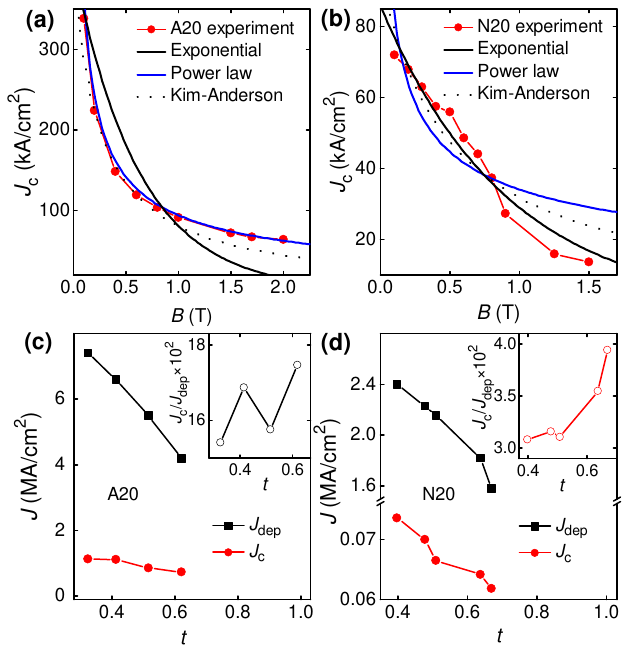}
  \caption{
  Experimental dependences $J_\mathrm{c}(B)$ (symbols) for as-grown film A20 at $T=3$\,K (a) and annealed film N20 at $T=2.5$\,K (b), together with fits to exponential, power-law, and Kim–Anderson models. Temperature dependence of the critical ($J_\mathrm{c}$) and pair-breaking ($J_\mathrm{dep}$) currents at zero magnetic field for the as-grown (c) and annealed (d) films. Insets in (c) and (d) show the ratio of $J_\text{c}$ to $J_\text{dep}$ at different normalized temperatures for the same films.
  \label{Jc}
  }
  \end{figure}
  
The critical current was deduced from the $I$-$V$ curves by using a voltage criterion of $1$\,mV.
With increasing magnetic field, the critical current density $J_\mathrm{c}(B)$ decreases in both the as-grown and annealed samples, as shown in Figs.~\ref{Jc}(a) and (b). $J_\mathrm{c}(T)$ decreases monotonically with increase of temperature [Figs.~\ref{Jc}(c) and (d)], as expected. In the low-field regime, the critical current is expected to be influenced by the edge barrier for vortex entry, with the edge-barrier suppression field given by~\cite{plourde2001influence,maksimova1998mixed}
$B_{\mathrm{stop}} = \Phi_0/[2\sqrt{3}\pi\xi(T) W]$,
where $W$ is the strip width. Note that our measurements were performed at magnetic fields well above $B_\mathrm{stop}$, which is of the order of a few mT in the studied samples. Accordingly, a vortex-free regime is not accessible in our measurements. In the mixed state, $J_\mathrm{c}$ is governed by vortex pinning \cite{ruiz2026critical,Dob20nac}, and our experimental data were compared with different models, among which the best possible fits are shown in Fig.~\ref{Jc}(a) and (b). Specifically, the power-law expression $J_\mathrm{c}(B)= A/B^{n}$ implies collective vortex pinning \cite{ruiz2026critical,badia2012electromagnetics}
while the exponential dependence $J_\mathrm{c}(B) = A \exp(-B/B_0)$, where $A$ is a constant and $B_0$ the characteristic field, is expected for highly disordered systems~\cite{riva2021wide}. The Kim-Anderson model $J_\mathrm{c}(B) = J_\mathrm{c0}/(1 + B/B_0)$ is applicable for pinning by strong, localized pinning centers \cite{anderson1964hard}.

For as-grown film A20 [Fig.~\ref{Jc}(a)], the $J_\mathrm{c}(B)$ data are best described by a power-law model with exponent $n\simeq 0.6$, suggesting collective vortex pinning in a relatively homogeneous SC network~\cite{blatter1994vortices}. The slow decay of $J_\mathrm{c}$ reflects robust pinning and good intergranular connectivity. In contrast, annealed film N20 [Fig.~\ref{Jc}(b)] exhibits a much faster suppression of $J_\mathrm{c}(B)$ at fields above $0.7$\,T. The exponential model provides the best fit, with parameters $A=86$\,kA/cm$^2$ and $B_0=0.92$\,T. This behavior signals reduced effective pinning and weakened intergranular coupling~\cite{anderson1964hard,larbalestier2001high}. In this regime, vortices are expected to move more easily along oxidized grain boundaries. Thus, the contrasting $J_\mathrm{c}(B)$ behavior highlights the impact of annealing-induced microstructural changes. While as-grown film A20 remains governed by collective pinning, annealed film N20 crosses over to a granularity-dominated regime. We assume that, in this regime, current transport is further limited by Josephson-like coupling across grain boundaries.

At $H=0.1$\,T and $3$\,K, $J_\mathrm{c}$ for samples A20 and N20 is approximately $350\,\mathrm{kA/cm^2}$ and $75\,\mathrm{kA/cm^2}$, respectively. The significantly larger $J_\mathrm{c}$ for as-grown sample A20 is attributed to a more uniform current distribution arising from stronger intergranular coupling. A comparison between the as-grown and annealed films therefore indicates that, although annealing increases the average crystallite size, it degrades the global current-carrying capability. This reduction is attributed to the formation of oxidized grain boundaries and weak links, which limit the macroscopic critical current. 

The maximal dissipation-free current a superconductor can carry is limited by the pair-breaking current, whose temperature dependence is given by
$I_\mathrm{dep}(T)=
I_\mathrm{dep}(0)[1-(T/T_\mathrm{c})^2]^{3/2}$ with
$I_\mathrm{dep}(0)=
0.74\,W[\Delta(0)]^{3/2}/
(eR_{\square}\hbar D\sqrt{1+W/(\pi\Lambda)})
$ in the dirty limit \cite{kuprianov1988influence}. Here, $\Delta(0)$ is the SC gap at zero temperature. For film A20, the ratio $J_\mathrm{c}/J_\mathrm{dep}$ at $2.5$\,K is about $15\%$, see the inset in Fig.~\ref{Jc}(c). Using the BCS ratio $\Delta(0) \approx 1.76\,k_\mathrm{B}T_\mathrm{c}$, we estimate for sample N20 in zero magnetic field a depairing current density of $J_\mathrm{dep}(2.5\,\mathrm{K}) \approx 2.8$\,MA/cm$^2$, while the experimentally measured zero-field critical current density is $J_\mathrm{c}(2.5\,\mathrm{K}) \approx 80$\,kA/cm$^2$. Thus, the ratio $J_\mathrm{c}/J_\mathrm{dep}$ for sample N20 is approximately $3.5\%$ over the investigated temperature range, see the inset in Fig.~\ref{Jc}(d). This ratio indicates that $I_\mathrm{c}$ is not limited by the intrinsic pair-breaking current, but instead by extrinsic factors such as granularity, weak-link behavior, and thermally-assisted vortex motion\,\cite{kupriyanov1980temperature,romijn1982critical,blatter1994vortices,larbalestier2001high}.

\section{Discussion}
\label{sec:discussion}
\subsection{Experimental findings}
First, we summarize the main experimental findings. Compared to the as-grown films, the annealed samples exhibit a coarser grain morphology and a significantly higher resistivity. At the same time, the SC transition temperature is only slightly reduced, while the transition becomes narrower than in the as-grown state, indicating improved homogeneity and enhanced SC properties within individual grains. The magnetic-field dependence of the critical current shows a crossover from a power-law behavior in the as-grown state to an exponential dependence after annealing, accompanied by a reduction of $J_\mathrm{c}$ by more than an order of magnitude. Furthermore, while the as-grown samples display single-step transitions to the normal state, the annealed films exhibit pronounced voltage kinks in their $I$-$V$ characteristics, suggesting the formation of a disordered network of weakly coupled grains. Taken together, these observations indicate that annealing enhances superconductivity within the grains themselves, while the grain boundaries act as regions of suppressed order, forming a branched network that governs the global transport properties.

\begin{figure*}[t]
    \centering
    \includegraphics[width=1\linewidth]{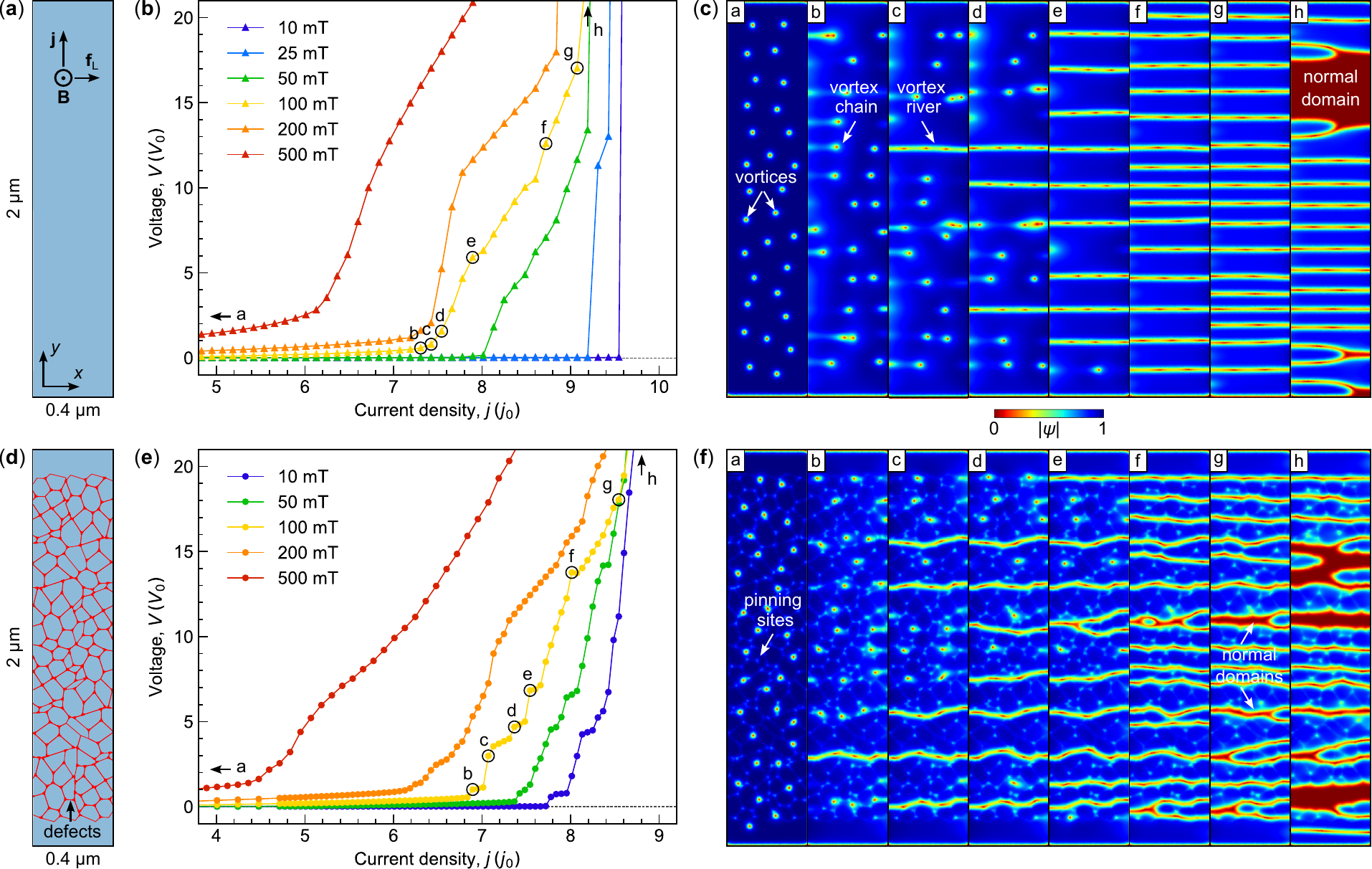}
    \caption{
    TDGL modeling results for the as-grown (a-c) and annealed films (d-f). (a,\,d) Geometries of the SC strips, indicating the directions of the applied magnetic field and transport current, with a grain-boundary mesh with suppressed order parameter shown in (d). Calculated $I$-$V$ curves for the as-grown (b) and annealed (e) samples. Labeled points (a-h) denote representative dynamical states, with snapshots of the spatial maps of the order-parameter magnitude $|\psi|$ shown in panels (c) and (f), respectively. The current density $j$ is expressed in units of $j_0 = 4\times 10^{-2}\,\xi B_\mathrm{c2}/(\mu_0 \lambda^2)$, where $B_\mathrm{c2} = \Phi_0/(2\pi\xi^2)$ is the upper critical field, and the voltage $V$ is in units of $V_0 = \xi j_0/\sigma$, with $\sigma$ being the normal-state conductivity.
    }
    \label{fig:TDGL_results}
\end{figure*}

The presence of such a complex pinning landscape in the annealed sample implies that vortex dynamics do not occur uniformly across the entire sample but are instead concentrated along channels with a suppressed SC order parameter. Consequently, the standard relation $v^\ast = V^{\ast}/(BL)$ which is justified for global FFI models is no longer applicable. Indeed, if one applies this relation, the deduced $v^\ast$ will range from approximately $0.1$ to $0.7$\,km/s for the first voltage jump, referred to as kink 0, and---if formally interpreted as vortex velocities---would reach up to $\sim 20$\,km/s for higher-order voltage kinks. In principle, in thin films, such high vortex velocities are physically plausible, with $v^\ast \sim 40$\,km/s in Pb, $\sim 20$\,km/s in MoSi~\cite{budinska2022rising}, and $\sim 15$\,km/s in NbC~\cite{Dob20nac}. However, the expression $v^\ast = V^{\ast}/(BL)$ is not valid when the resistive transition is governed by mechanisms other than vortex motion, including phase slips of the SC order parameter~\cite{Siv03prl} and the nucleation of N domains~\cite{Bez84ltp}. Furthermore, even when the underlying mechanism is related to vortex motion, the relation $v^\ast = V^{\ast}/(BL)$ becomes inapplicable if the actual number of vortices in the sample deviates from $m\Phi_0 = BS$, where $m=1,2,\dots$ and $S$ is the sample area~\cite{Dob20nac,bevz2023vortex}, if vortex velocities vary significantly across the sample~\cite{Gri15prb,Sil12njp,Ada15prb}, or if FFI emerges only in localized regions~\cite{bezuglyj2019local}. 

In general, in spatially uniform constrictions, phase slips lead to voltage states whose extrapolation to the current axis defines a joint crossing point corresponding to the excess current~\cite{Zol13ltp}. In the mixed state, channels of fast-moving vortices may display behavior analogous to phase-slip lines~\cite{Vod19sst}, forming vortex rivers~\cite{Sil10prl}---regions with a locally suppressed SC order parameter~\cite{Emb17nac}. A fast-vortex-dynamics-induced phase-slip regime can occur when $J^\ast < J_\mathrm{eq}$, whereas vortex rivers will develop into N domains for $J^\ast > J_\mathrm{eq}$. Accordingly, since $J^\ast > J_\mathrm{eq}$ for our samples, the transition to the highly resistive state should involve the formation of N domains. 

We note that the low-resistive state is not destroyed instantaneously throughout the entire volume of the annealed samples, as evidenced by the appearance of voltage kinks over a finite range of transport currents. This indicates that the rate of heat removal remains sufficiently high compared to the power dissipated at the characteristic currents $I^\ast$, so that the system does not immediately collapse into the normal state upon the appearance of N domains, but instead allows a sequence of distinct states to develop during the resistive transition. Microscopically, these states can be associated with variations in the number, spatial distribution, and size of the N domains. To shed light on the spatiotemporal evolution of the SC order parameter in the samples during the resistive transition, we performed modeling based on a numerical solution of the TDGL equation. The results of these simulations are discussed next.

\subsection{TDGL modeling}
Simulations of the vortex dynamics and the associated $I$-$V$ curves were performed on the basis of the generalized TDGL equation \cite{Kra78prl, Bis23cpc}, as detailed in the Appendix. 

The simulation results are shown in Fig.~\ref{fig:TDGL_results}(a–c) for the as-grown film and in Fig.~\ref{fig:TDGL_results}(d–f) for the annealed film. Figures~\ref{fig:TDGL_results}(a) and (d) show the simulated geometries. In Fig.~\ref{fig:TDGL_results}(d), a network of linearly extended regions with a suppressed SC order parameter is used to model the grain boundaries as a disordered network in the annealed film.

Figure~\ref{fig:TDGL_results}(b) displays the calculated $I$-$V$ curves for the as-grown film in magnetic fields ranging from 10 to 500 mT. For 10\,mT and 25\,mT, the simulated $I$-$V$ curves exhibit a zero-voltage regime followed by abrupt jumps to the normally conducting state, just as observed in the experiment. At higher magnetic fields, low-resistive sections appear at low currents. Qualitatively, these resistive regimes are observed in the experimental $I$-$V$ curves in Fig.~\ref{I-V}(a) for magnetic fields above $400$\,mT. The quantitative discrepancy between simulation and  experiment is attributed to the smaller dimensions of the constriction and the neglect of volume pinning in the simulations.

We now consider in detail the $I$-$V$ curve for $100$\,mT in Fig.~\ref{fig:TDGL_results}(b), which is representative for moderately high vortex densities. At low currents [point a in Fig.~\ref{fig:TDGL_results}(b,\,c)] the vortices are immobile. Their motion is prevented by the edge barrier~\cite{Vod19sst} up to $j/j_0 \lesssim 7$, above which the edge barrier is suppressed by the transport current and the vortices begin to move, forming vortex chains [point b in Fig.~\ref{fig:TDGL_results}(b,\,c)]. The recovery of the SC order parameter behind the moving vortices is delayed, causing the vortex chains to evolve into vortex rivers [points c and d in Fig.~\ref{fig:TDGL_results}(b,\,c)]. With a further increase of $j/j_0 \gtrsim 7.8$ the number of vortex rivers increases [points e, f, and g in Fig.~\ref{fig:TDGL_results}(b,\,c)] and the distance between them decreases. When the distance between neighboring vortex rivers becomes too small to dissipate the generated heat, they merge and form an N domain which grows until the entire strip transitions to the normal conducting state. Accounting for the finite rate of heat removal to the substrate complicates TDGL modeling~\cite{Dob20nac}; we have checked that its including (not shown) eliminates the low-resistive regimes between b and g in Fig.~\ref{fig:TDGL_results}(b,\,c) and shifts the transition to the normal state towards lower currents.

Figures~\ref{fig:TDGL_results}(e) and (f) show the $I$-$V$ curves and snapshots of the SC order parameter magnitude, $|\psi(x,y)|$, at selected points for the annealed film at $100$\,mT. Below the depinning threshold, vortices remain pinned to structural defects, producing a zero-voltage response [point a in Fig.~\ref{fig:TDGL_results}(e,\,f)]. A nonzero voltage appears when vortices depin and follow curved paths along grain boundaries, unlike the straight chains in the as-grown film. The higher resistivity at these boundaries induces stronger local overheating \cite{bezuglyj2019local}, transforming fast-moving vortex chains into vortex rivers, which subsequently evolve into N domains [points e and f in Fig.~\ref{fig:TDGL_results}(e,f)]. The number of domains reflects the available paths where superconductivity is sufficiently suppressed, while the stronger order parameter and lower resistivity within the grains limit domain growth. As the transport current increases further, the N domains expand [points g and h in Fig.~\ref{fig:TDGL_results}(e,f)] until the entire film becomes normal conducting.

In this way, the TDGL simulation results indicate that the observed voltage kinks arise from the formation of several N domains, rather than from a single abrupt transition to a highly dissipative state. This mechanism is consistent with a heating-dominated FFI model~\cite{bezuglyj2019local}, but is spatially non-uniform due to the coarse grains in the annealed film. Notably, the sequence of dynamic states in the annealed sample differs from other scenarios, including the formation of temperature gradients in microstructurally uniform strips with strong edge barriers~\cite{Dob20nac}, overheating of central regions in strips with pronounced current crowding~\cite{Emb17nac}, attraction of fast vortex trajectories to periodically positioned pinning sites~\cite{Ada15prb}, and the nucleation of FFIs in one or more straight channels~\cite{bezuglyj2019local}. 

\section{Conclusion}
We have studied the electrical resistance and superconducting properties of as-grown and annealed NbRe thin films. Annealing has been revealed to coarsen the grains, weaken the intergranular coupling, and reduce the effective pinning landscape. As a result, vortex motion in the annealed films becomes highly inhomogeneous, leading to lower critical currents and reduced flux-flow instability velocities. The annealed films exhibit voltage kinks in their $I$-$V$ curves, reflecting the nucleation and growth of normal domains. Time-dependent Ginzburg–Landau simulations reproduce the major features of the $I$-$V$ curves and show that fast vortex motion and domain formation are guided by extended grain boundaries. Overall, the annealed films behave as microstructurally engineered superconducting networks, where vortices depin easily but move through a strongly nonuniform environment, resulting in weak pinning and enhanced local overheating. From an application perspective, the localized heating and successive voltage kinks may offer potential for superconducting sensors and threshold detectors, where controlled discrete resistive states can be exploited for signal transduction~\cite{semenov2001quantum,Buh15nac}. In this way, the oxidation introduced during annealing effectively tailors the superconducting network, enabling vortex dynamics and dissipation patterns that are difficult to achieve in uniform superconductors.

\section*{Acknowledgements}
The work of Z.M.K. and A.P. was funded by the Deutsche Forschungsgemeinschaft (DFG, German Research Foundation) under Germany's Excellence Strategy -- EXC-2123 QuantumFrontiers -- 390837967, projects FF-145 (Super3D) and Q-53 (Opto-Fluxonics). Z.M.K. and A.P. gratefully acknowledge the support of the Braunschweig International Graduate School of Metrology (B-IGSM). A.P. gratefully acknowledges the use of the CryoCore and CryoCube simulation workstations at CryoQuant/TU Braunschweig.

\section*{Appendix: TDGL modeling}
\label{sec:TDGL}
The spatiotemporal evolution of the SC order parameter was modeled via a numerical solution of the TDGL equation. Although ongoing hard point contact spectroscopy (PCS) studies suggest possible multiple SC gaps in NbRe single crystals \cite{Cirillo2015}, their presence in thin films has yet to be conclusively established. Accordingly, we adopted a strong interband-coupling approximation in the TDGL modeling \cite{Dol08prb, Gri16prb}, in which rapid interband Cooper-pair tunneling locks the order parameters of all bands. This renders the system effectively single-band with respect to vortex dynamics, permitting a description within a single-band TDGL framework.

In the 2D geometry, the generalized TDGL\,\cite{Kra78prl, Bis23cpc} reads
\begin{equation}
    \begin{split}
    \frac{u}{\sqrt{1+\gamma^2|\psi|^2}}&\left(\frac{\partial}{\partial t}+i\varphi+\frac{\gamma^2}{2}\frac{\partial |\psi|^2}{\partial t}\right)\psi\\
    &= (\nabla-i\mathbf{A})^2\psi + (\epsilon-|\psi|^2)\psi
    ,
    \end{split}
    \label{eq:tdgl_psi}
\end{equation}
where $\psi(\mathbf{r},t) = |\psi| e^{i \theta}$ is the SC order parameter with phase $\theta$, $u=\pi^4/14\zeta(3)\approx5.79$ is the ratio of the relaxation times for the amplitude and phase of the order parameter in the dirty limit, with $\zeta(x)$ denoting the Riemann zeta function. The parameter $\gamma=2\tau_E\Delta_0/\hbar$ involves the inelastic electron-phonon scattering time $\tau_\mathrm{E}$ and the zero-field SC gap $\Delta_0$. $\varphi(\mathbf{r}, t)$ is the electric scalar potential and $\mathbf{A}$ the vector potential. The real-valued parameter $\epsilon(\mathbf{r}) \in [-1, 1]$~\cite{KosPRB16} describes the suppression of the  order parameter along the grain boundaries. 

The total current density $\mathbf{j}_\mathrm{ext}$ is composed of the supercurrent density $\mathbf{j}_\mathrm{s} = \mathrm{Im}[\psi^\ast(\nabla-i\mathbf{A})\psi]$ and the normal current density $\mathbf{j}_n = -\nabla\varphi - \partial \mathbf{A}/\partial t$. The  electric potential $\varphi(\mathbf{r}, t)$ satisfies the Poisson equation 
\begin{equation}
    \nabla^2\varphi = \nabla\cdot\mathrm{Im}[\psi^\ast(\nabla-i\mathbf{A})\psi] - \nabla \cdot \dfrac{\partial \mathbf{A}}{\partial t},
    \label{eq:tdgl_poisson}
\end{equation}
which follows from the continuity condition for the total current density, $\nabla\cdot (\mathbf{j}_\mathrm{s} + \mathbf{j}_\mathrm{n}) = 0$.

At the edges where vortices enter or exit the strip, the boundary conditions are 
\(\mathbf{n}\cdot(\nabla-i\mathbf{A})\psi = 0\) and \(\mathbf{n}\cdot\nabla\varphi = 0\), 
where \(\mathbf{n}\) is a unit vector normal to the interface. 
At the N/S interfaces, through which \(j_\mathrm{ext}\) is applied, 
the boundary conditions are \(\psi = 0\) and \(\mathbf{n}\cdot\nabla\varphi = j_\mathrm{ext}\).

In Eqs.\,\eqref{eq:tdgl_psi} and \eqref{eq:tdgl_poisson}, all quantities are dimensionless. The order-parameter magnitude is scaled to the SC charge-carrier density, $|\psi|^2 = n_\mathrm{s}$. Lengths are in units of $\xi$, and time $t$ is scaled by $\tau_0 = \mu_0 \sigma \lambda^2$, where $\sigma$ denotes the normal-state conductivity and $\lambda$ the London penetration depth. The vector potential $\mathbf{A}$ is scaled by $\xi B_\mathrm{c2}$, where $B_\mathrm{c2} = \Phi_0 / (2\pi \xi^{2})$ is the upper critical field and $\Phi_0 = h / (2e)$ is the magnetic flux quantum. The total current density $j_\mathrm{ext}$ and electric potential $\varphi$ are scaled by $4 \xi B_\mathrm{c2} / (\mu_0 \lambda^2)$ and $4 \xi^2 B_\mathrm{c2} / (\sigma \mu_0 \lambda^2)$, respectively. 

The TDGL simulations were carried out for $2\times 0.4\,\mu$m (length$\times$width) rectangular strips in magnetic fields applied normally to the film plane. The maximum mesh size was set to $\xi/2$. The material parameters are summarized in Table~\ref{tab:tdgl_material_params}. 

\begin{table}[h!]
\caption{Material parameters used in the simulations.}
\label{tab:tdgl_material_params}
\small
\begin{center}
\begin{tabular}{lcc}
 \hline
 \textbf{Parameter} & \textbf{Denotation} & \textbf{Value}\\
 \hline
 Clean-limit coherence length & $\xi_0$ & 18 nm \\
 Electron mean free path & $l$ & 3 nm \\
 Coherence length & $\xi(0) = 0.85 \, \sqrt{\xi_0 l}$ & 6.3 nm \\
 London penetration depth & $\lambda(0)$ & 760 nm \\
 Inelastic scattering coefficient & $\gamma$ & 7 \\
 Temperature & $T$ & 3.2 K \\
 Critical temperature & $T_\mathrm{c}$ & 6.3 K \\
 Coherence length & $\xi=\frac{\xi(0)}{\sqrt{1-T/T_\mathrm{c}}}$ & 9\,nm \\
 Penetration depth & $\lambda= \frac{0.615 \, \lambda(0)\sqrt{\xi_0}}{\sqrt{l (1-T/T_\mathrm{c})}}$ & 1654\,nm \\
 \hline
\end{tabular}
\end{center}
\end{table}

The as-grown film was modeled as a spatially uniform strip, since the coherence length $\xi(0) \approx 6$\,nm exceeds the average grain size of about $2$\,nm. In contrast, the annealed film was modeled as a strip containing a network of regions, each with a width of $5–15\xi$, where the order parameter is suppressed, as illustrated in Fig.~\ref{fig:TDGL_results}(d). The connecting segments of the network were assigned a width of $\xi/2$ and $\epsilon(\mathbf{r})=0$, while the nodes at their intersections were assigned a width of $\xi$ and $\epsilon(\mathbf{r})=-1$.

\bibliography{references}

@article{Cirillo2016,
  author  = {C. Cirillo and G. Carapella and M. Salvato and R. Arpaia and M. Caputo and C. Attanasio},
  title   = {Superconducting properties of noncentrosymmetric {Nb}$_{0.18}${Re}$_{0.82}$ thin films probed by transport and tunneling experiments},
  journal = {Phys. Rev. B},
  volume  = {94},
  pages   = {104512},
  year    = {2016},
  doi     = {10.1103/PhysRevB.94.104512}
}

@article{Cirillo2015,
  author  = {C. Cirillo and R. Fittipaldi and M. Smidman and G. Carapella and C. Attanasio and A. Vecchione and R. P. Singh and M. R. Lees and G. Balakrishnan and M. Cuoco},
  title   = {Evidence of double‑gap superconductivity in noncentrosymmetric {Nb}$_{0.18}${Re}$_{0.82}$ single crystals},
  journal = {Phys. Rev. B},
  volume  = {91},
  pages   = {134508},
  year    = {2015},
  doi     = {10.1103/PhysRevB.91.134508}
}

@article{Karki2011,
  author  = {A. B. Karki and Y. M. Xiong and N. Haldolaarachchige and S. Stadler and I. Vekhter and P. W. Adams and D. P. Young and W. A. Phelan and J. Y. Chan},
  title   = {Physical properties of the noncentrosymmetric superconductor {Nb}$_{0.18}${Re}$_{0.82}$},
  journal = {Phys. Rev. B},
  volume  = {83},
  pages   = {144525},
  year    = {2011},
  doi     = {10.1103/PhysRevB.83.144525}
}

@article{Chen2013,
  title={{BCS}-like superconductivity in the noncentrosymmetric compounds {Nb}$_x${Re}$_{1-x}$ (0.13 $\leq x \leq$ 0.38)},
  author={J. Chen and L. Jiao and J. L. Zhang and Y. Chen and L. Yang and M. Nicklas and F. Steglich and H. Q. Yuan},
  journal={Phys. Rev. B},
  volume={88},
  pages={144510},
  year={2013},
  doi={10.1103/PhysRevB.88.144510}
}

@ARTICLE{Buh15nac,
  author = {Buh, J. and Kabanov, V. and Baranov, V. and Mrzel, A. and Kovic,
	A. and Mihailovic, D.},
  title = {Control of switching between metastable superconducting states in
	{d-MoN} nanowires},
  journal = {Nat. Commun.},
  year = {2015},
  volume = {6},
  pages = {10250},
  day = {21},
  publisher = {The Author(s) SN -},
  url = {http://dx.doi.org/10.1038/ncomms10250}
}

@ARTICLE{Vod19sst,
  author = {Vodolazov, D. Yu.},
  title = {Flux-flow instability in a strongly disordered superconducting strip
	with an edge barrier for vortex entry},
  journal = {Supercond. Sci. Technol.},
  year = {2019},
  volume = {32},
  pages = {115013},
  number = {11},
  month = {oct},
  doi = {10.1088/1361-6668/ab4168},
  publisher = {{IOP} Publishing},
  url = {https://doi.org/10.1088/1361-6668/ab4168}
}

@ARTICLE{Ada15prb,
  author = {Adami, O.-A. and Jelic, Z. L. and Xue, C. and Abdel-Hafiez, M. and
	Hackens, B. and Moshchalkov, V. V. and Milosevic, M. V. and Van de
	Vondel, J. and Silhanek, A. V.},
  title = {Onset, evolution, and magnetic braking of vortex lattice instabilities
	in nanostructured superconducting films},
  journal = {Phys. Rev. B},
  year = {2015},
  volume = {92},
  pages = {134506},
  month = {Oct},
  doi = {10.1103/PhysRevB.92.134506},
  issue = {13},
  numpages = {9},
  publisher = {American Physical Society},
  url = {https://link.aps.org/doi/10.1103/PhysRevB.92.134506}
}

@article{cirillo2022polycrystalline,
  title={Polycrystalline {NbRe} superconducting films deposited by direct current magnetron sputtering},
  author={C. Cirillo and M. Caputo and G. Divitini and J. W. A. Robinson and C. Attanasio},
  journal={Thin Solid Films},
  volume={758},
  pages={139450},
  year={2022},
  doi={10.1016/j.tsf.2022.139450}
}

@article{makhdoumi2024effect,
  title={Effect of thermal annealing on the average and local structure of superconducting polycrystalline {NbRe} films},
  author={{Z. Makhdoumi Kakhaki} and A. Martinelli and F. Avitabile and A. Di Bernardo and C. Attanasio and C. Cirillo},
  journal={Supercond. Sci. Technol.},
  volume={37},
  number={12},
  pages={125002},
  year={2024},
  doi={10.1088/1361-6668/ad8122}
}

@article{caputo2017nbre,
  author  = {M. Caputo and C. Cirillo and C. Attanasio},
  title   = {{NbRe} as candidate material for fast single photon detection},
  journal = {Appl. Phys. Lett.},
  volume  = {111},
  pages   = {192601},
  year    = {2017},
  doi     = {10.1063/1.4997675}
}

@article{larkin1975nonlinear,
  title={Nonlinear conductivity of superconductors in the mixed state},
  author={A. I. Larkin and Yu. N. Ovchinnikov},
  journal={Sov. Phys. JETP},
  volume={41},
  number={5},
  pages={960},
  year={1975}
}

@article{bezuglyj1992effect,
  author  = {A. I. Bezuglyj and V. A. Shklovskij},
  title   = {Effect of self‑heating on flux flow instability in a superconductor near ${T}_{\mathrm{c}}$},
  journal = {Physica C},
  volume  = {202},
  pages   = {234--242},
  year    = {1992},
  doi     = {10.1016/0921-4534(92)90165-9}
}

@article{bezuglyj2019local,
  author  = {A. I. Bezuglyj and V. A. Shklovskij and R. V. Vovk and V. M. Bevz and M. Huth and O. V. Dobrovolskiy},
  title   = {Local flux‑flow instability in superconducting films near ${T_c}$},
  journal = {Phys. Rev. B},
  volume  = {99},
  pages   = {174518},
  year    = {2019},
  doi     = {10.1103/PhysRevB.99.174518}
}

@article{bevz2023vortex,
  author  = {V. M. Bevz and M. Yu. Mikhailov and B. Budinsk{\'a} and S. Lamb‑Camarena and S. O. Shpilinska and A. V. Chumak and M. Urb{\'a}nek and M. Arndt and W. Lang and O. V. Dobrovolskiy},
  title   = {Vortex counting and velocimetry for slitted superconducting thin strips},
  journal = {Phys. Rev. Appl.},
  volume  = {19},
  pages   = {034098},
  year    = {2023},
  doi     = {10.1103/PhysRevAppl.19.034098}
}

@article{plourde2001influence,
  author  = {B. L. T. Plourde and D. J. Van Harlingen and D. Yu. Vodolazov and R. Besseling and M. B. S. Hesselberth and P. H. Kes},
  title   = {Influence of edge barriers on vortex dynamics in thin weak‑pinning superconducting strips},
  journal = {Phys. Rev. B},
  volume  = {64},
  pages   = {014503},
  year    = {2001},
  doi     = {10.1103/PhysRevB.64.014503}
}

@article{budinska2022rising,
  author  = {B. Budinsk{\'a} and B. Aichner and D. Yu. Vodolazov and M. Yu. Mikhailov and F. Porrati and M. Huth and A. V. Chumak and W. Lang and O. V. Dobrovolskiy},
  title   = {Rising speed limits for fluxons via edge‑quality improvement in wide {MoSi} thin films},
  journal = {Phys. Rev. Appl.},
  volume  = {17},
  pages   = {034072},
  year    = {2022},
  doi     = {10.1103/PhysRevApplied.17.034072}
}

@article{maksimova1998mixed,
  title={Mixed state and critical current in narrow semiconducting films},
  author={Maksimova, G. M.},
  journal={Phys. Solid State},
  volume={40},
  number={10},
  pages={1607--1610},
  year={1998},
  doi= {10.1134/1.1130618}
}

@article{ruiz2026critical,
  title={Critical current density in advanced superconductors},
  author={H. S. Ruiz and J. H{\"a}nisch and M. Polichetti and A. Galluzzi and L. Gozzelino and D. Torsello and S. Milo{\v{s}}evi{\'c}-Govedarovi{\'c} and J. Grbovi{\'c}-Novakovi{\'c} and O. V. Dobrovolskiy and W. Lang and others},
  journal={Prog. Mater. Sci.},
  volume={155},
  pages={101492},
  year={2026},
  doi={10.1016/j.pmatsci.2025.101492}
}

@ARTICLE{Gur84spu,
  author = {Gurevich, A. V. and Mints, R. G.},
  journal = {Sov. Phys. Usp.},
  year = {1984},
  volume = {27},
  pages = {19}
}

@ARTICLE{Bez84ltp,
  author = {Bezuglyj, A. I. and Shklovskij, V. A.},
  title = {Thermal domains in inhomogeneous current-carrying superconductors.
	Current-voltage characteristics and dynamics of domain formation
	after current jumps},
  journal = {J. Low Temp. Phys.},
  year = {1984},
  volume = {57},
  pages = {227--247},
  number = {3},
  month = {Nov},
  day = {01},
  doi = {10.1007/BF00681190},
  issn = {1573-7357},
  url = {https://doi.org/10.1007/BF00681190}
}

@ARTICLE{Gri15prb,
  author = {Grimaldi, G. and Leo, A. and Sabatino, P. and Carapella, G. and Nigro,
	A. and Pace, S. and Moshchalkov, V. V. and Silhanek, A. V.},
  title = {Speed limit to the {Abrikosov} lattice in mesoscopic superconductors},
  journal = {Phys. Rev. B},
  year = {2015},
  volume = {92},
  pages = {024513},
  month = {Jul},
  doi = {10.1103/PhysRevB.92.024513},
  issue = {2},
  numpages = {6},
  publisher = {American Physical Society},
  url = {http://link.aps.org/doi/10.1103/PhysRevB.92.024513}
}

@ARTICLE{Zol13ltp,
  author = {I. V. Zolochevskii},
  title = {Stimulation of superconductivity by microwave radiation in wide tin
	films},
  journal = {Low Temp. Phys.},
  year = {2013},
  volume = {39},
  pages = {571},
  doi = {10.1063/1.4813655}
}

@article{anderson1964hard,
  author  = {P. W. Anderson and Y. B. Kim},
  title   = {Hard superconductivity: theory of the motion of {Abrikosov} flux lines},
  journal = {Rev. Mod. Phys.},
  volume  = {36},
  pages   = {39--43},
  year    = {1964},
  doi     = {10.1103/RevModPhys.36.39}
}

@article{badia2012electromagnetics,
  author  = {A. Bad{\'\i}a‑Maj{\'o}s and C. L{\'o}pez},
  title   = {Electromagnetics close beyond the critical state: thermodynamic prospect},
  journal = {Supercond. Sci. Technol.},
  volume  = {25},
  pages   = {104004},
  year    = {2012},
  doi     = {10.1088/0953-2048/25/10/104004}
}

@article{riva2021wide,
  author  = {N. Riva and F. Sirois and C. Lacroix and F. Pellerin and J. Giguere and F. Grilli and B. Dutoit},
  title   = {A wide range \(E\)--\(J\) constitutive law for simulating REBCO tapes above their critical current},
  journal = {Supercond. Sci. Technol.},
  volume  = {34},
  pages   = {115014},
  year    = {2021},
  doi     = {10.1088/1361-6668/ac2883}
}

@article{kuprianov1988influence,
  author  = {M. Yu. Kupriyanov and V. F. Lukichev},
  title   = {Influence of boundary transparency on the critical current of dirty S--S$'$--S structures},
  journal = {Zh. Eksp. Teor. Fiz.},
  volume  = {94},
  pages   = {139},
  year    = {1988}
}

@article{kupriyanov1980temperature,
  author  = {M. Yu. Kupriyanov and V. F. Lukichev},
  title   = {Temperature dependence of pair‑breaking current in superconductors},
  journal = {Sov. J. Low Temp. Phys.},
  volume  = {6},
  pages   = {210--214},
  year    = {1980},
  doi     = {10.1063/10.0030039}
}

@article{romijn1982critical,
  author  = {J. Romijn and T. M. Klapwijk and M. J. Renne and J. E. Mooij},
  title   = {Critical pair‑breaking current in superconducting aluminum strips far below ${T}_c$},
  journal = {Phys. Rev. B},
  volume  = {26},
  pages   = {3648--3655},
  year    = {1982},
  doi     = {10.1103/PhysRevB.26.3648}
}

@article{cirillo2020superconducting,
  author  = {C. Cirillo and J. Chang and M. Caputo and J. W. N. Los and S. Dorenbos and I. Esmaeil Zadeh and C. Attanasio},
  title   = {Superconducting nanowire single photon detectors based on disordered {NbRe} films},
  journal = {Appl. Phys. Lett.},
  volume  = {117},
  pages   = {172602},
  year    = {2020},
  doi     = {10.1063/5.0021487}
}

@article{ejrnaes2022single,
  author  = {M. Ejrnaes and C. Cirillo and D. Salvoni and F. Chianese and C. Bruscino and P. Ercolano and A. Cassinese and C. Attanasio and G. P. Pepe and L. Parlato},
  title   = {Single photon detection in {NbRe} superconducting microstrips},
  journal = {Appl. Phys. Lett.},
  volume  = {121},
  pages   = {265303},
  year    = {2022},
  doi     = {10.1063/5.0131336}
}

@article{cirillo2024single,
  author  = {C. Cirillo and M. Ejrnaes and P. Ercolano and C. Bruscino and A. Cassinese and D. Salvoni and C. Attanasio and G. P. Pepe and L. Parlato},
  title   = {Single photon detection up to 2 \(\mu\)m in pair of parallel microstrips based on {NbRe} ultrathin films},
  journal = {Sci. Rep.},
  volume  = {14},
  pages   = {20345},
  year    = {2024},
  doi     = {10.1038/s41598-024-66991-1}
}

@inbook{Dob24inb,
  author    = {O. V. Dobrovolskiy},
  title     = {Fast dynamics of vortices in superconductors},
  booktitle = {Encyclopedia of Condensed Matter Physics},
  edition   = {2},
  editor    = {V. M. Fomin},
  chapter   = {9},
  pages     = {735},
  publisher = {Elsevier},
  year      = {2024},
  doi       = {10.1016/B978-0-323-90800-9.00015-9}
}

@article{koch2024gate,
  author  = {J. Koch and C. Cirillo and S. Battisti and L. Ruf and {Z. Makhdoumi Kakhaki} and A. Paghi and A. Gulian and S. Teknowijoyo and G. De Simoni and F. Giazotto and C. Attanasio and E. Scheer and A. Di Bernardo},
  title   = {Gate‑controlled supercurrent effect in dry‑etched {Dayem} bridges of non‑centrosymmetric niobium rhenium},
  journal = {Nano Res.},
  volume  = {17},
  pages   = {6575},
  year    = {2024},
  doi     = {10.1007/s12274-024-6576-7}
}

@article{colangelo2025unveiling,
  title={Unveiling intrinsic triplet superconductivity in noncentrosymmetric {NbRe} through inverse spin-valve effects},
  author={F. Colangelo and M. Modestino and F. Avitabile and A. Galluzzi and {Z. Makhdoumi Kakhaki} and A. Kumar and J. Linder and M. Polichetti and C. Attanasio and C. Cirillo},
  journal={Phys. Rev. Lett.},
  volume={135},
  number={22},
  pages={226002},
  year={2025},
  doi={10.1103/PhysRevLett.135.226002}
}

@ARTICLE{Siv03prl,
  author = {Sivakov, A. G. and Glukhov, A. M. and Omelyanchouk, A. N. and Koval,
	Y. and M\"uller, P. and Ustinov, A. V.},
  title = {Josephson Behavior of Phase-Slip Lines in Wide Superconducting Strips},
  journal = {Phys. Rev. Lett.},
  year = {2003},
  volume = {91},
  pages = {267001},
  doi = {10.1103/PhysRevLett.91.267001},
  issue = {26},
  numpages = {4},
  publisher = {American Physical Society},
  url = {http://link.aps.org/doi/10.1103/PhysRevLett.91.267001}
}

@ARTICLE{Sil10prl,
  author = {Silhanek, A. V. and Milo\ifmmode \check{s}\else \v{s}\fi{}evi\ifmmode
	\acute{c}\else \'{c}\fi{}, M. V. and Kramer, R. B. G. and Berdiyorov,
	G. R. and Van de Vondel, J. and Luccas, R. F. and Puig, T. and Peeters,
	F. M. and Moshchalkov, V. V.},
  title = {Formation of Stripelike Flux Patterns Obtained by Freezing Kinematic
	Vortices in a Superconducting {Pb} Film},
  journal = {Phys. Rev. Lett.},
  year = {2010},
  volume = {104},
  pages = {017001},
  month = {Jan},
  doi = {10.1103/PhysRevLett.104.017001},
  issue = {1},
  numpages = {4},
  publisher = {American Physical Society},
  url = {https://link.aps.org/doi/10.1103/PhysRevLett.104.017001}
}

@ARTICLE{Sil12njp,
  author = {Silhanek, A. V. and Leo, A. and Grimaldi, G. and Berdiyorov, G. R.
	and Milosevic, M. V and Nigro, A. and Pace, S. and Verellen, N. and
	Gillijns, W. and Metlushko, V. and Ili\'c, B. and Zhu, X. and Moshchalkov,
	V. V.},
  title = {Influence of artificial pinning on vortex lattice instability in
	superconducting films},
  journal = {New J. Phys.},
  year = {2012},
  volume = {14},
  pages = {053006},
  number = {5},
  url = {http://stacks.iop.org/1367-2630/14/i=5/a=053006}
}

@ARTICLE{Emb17nac,
  author = {Embon, L. and Anahory, Y. and Jelic, Z. L. and Lachman, E. O. and
	Myasoedov, Y. and Huber, M. E. and Mikitik, G. P. and Silhanek, A.
	V. and Milosevic, M. V. and Gurevich, A. and Zeldov, E.},
  title = {Imaging of super-fast dynamics and flow instabilities of superconducting
	vortices},
  journal = {Nat. Commun.},
  year = {2017},
  volume = {8},
  pages = {85},
  doi = {10.1038/s41467-017-00089-3},
  issn = {2041-1723},
  url = {https://doi.org/10.1038/s41467-017-00089-3}
}

@ARTICLE{Dob20nac,
  author = {Dobrovolskiy, O. V. and Vodolazov, D. Yu and Porrati, F. and Sachser,
	R. and Bevz, V. M. and Mikhailov, M. Yu and Chumak, A. V. and Huth,
	M.},
  title = {Ultra-fast vortex motion in a direct-write {Nb-C} superconductor},
  journal = {Nat. Commun.},
  year = {2020},
  volume = {11},
  pages = {3291},
  number = {1},
  day = {03},
  doi = {10.1038/s41467-020-16987-y},
  issn = {2041-1723},
  url = {https://doi.org/10.1038/s41467-020-16987-y}
}

@article{kakhaki2024characterization,
  author  = {{Z. Makhdoumi Kakhaki} and A. Leo and A. Spuri and M. Ejrnaes and L. Parlato and G. P. Pepe and F. Avitabile and A. Di Bernardo and A. Nigro and C. Attanasio and C. Cirillo},
  title   = {Characterization of quasiparticle relaxation times in microstrips of {NbReN} for prospective applications in superconducting single‑photon detectors},
  journal = {Mater. Sci. Eng. B},
  volume  = {304},
  pages   = {117376},
  year    = {2024},
  doi     = {10.1016/j.mseb.2024.117376}
}

@article{blatter1994vortices,
  author  = {G. Blatter and M. V. Feigel’man and V. B. Geshkenbein and A. I. Larkin and V. M. Vinokur},
  title   = {Vortices in high‑temperature superconductors},
  journal = {Rev. Mod. Phys.},
  volume  = {66},
  number  = {4},
  pages   = {1125},
  year    = {1994},
  doi     = {10.1103/RevModPhys.66.1125}
}

@article{larbalestier2001high,
  author  = {D. Larbalestier and A. Gurevich and D.~M. Feldmann and A. Polyanskii},
  title   = {High‑${T}_c$ superconducting materials for electric power applications},
  journal = {Nature},
  volume  = {414},
  number  = {6861},
  pages   = {368--377},
  year    = {2001},
  doi     = {10.1038/35104654}
}

@article{Kra78prl,
title = {Theory of Dissipative Current-Carrying States in Superconducting Filaments},
author = {
Kramer, L. and 
Watts-Tobin, R. J.},
j1 = {PRL},
journal2 = {Physical Review Letters},
journal = {Phys. Rev. Lett.},
publisher = {American Physical Society},
volume = {40},
number = {15},
year = {1978},
month = {04},
pages = {1041--1044},
doi = {10.1103/PhysRevLett.40.1041},

date = {1978/04/10/}
}

@article{Bis23cpc,
title = {{pyTDGL}: {Time-dependent Ginzburg-Landau} in {Python}},
author = {Bishop-Van Horn, Logan},
journal2 = {Computer Physics Communications},
journal = {Comput. Phys. Commun.},
isbn = {0010-4655},
volume = {291},
year = {2023},
pages = {108799},
doi = {10.1016/j.cpc.2023.108799},

date = {2023/10/01/}
}

@article{KosPRB16,
  title = {Optimization of vortex pinning by nanoparticles using simulations of the time-dependent {Ginzburg-Landau} model},
  author = {Koshelev, A. E. and Sadovskyy, I. A. and Phillips, C. L. and Glatz, A.},
  journal = {Phys. Rev. B},
  volume = {93},
  issue = {6},
  pages = {060508},
  numpages = {5},
  year = {2016},
  month = {Feb},
  publisher = {American Physical Society},
  doi = {10.1103/PhysRevB.93.060508},
  url = {https://link.aps.org/doi/10.1103/PhysRevB.93.060508}
}

@article{semenov2001quantum,
  title={Quantum detection by current carrying superconducting film},
  author={A. D. Semenov and G. N. Gol’tsman and A. A. Korneev},
  journal={Physica C},
  volume={351},
  number={4},
  pages={349--356},
  year={2001},
  doi={10.1016/S0921-4534(00)01637-3}
}

@article{de2026fast,
  title={Fast vortex dynamics and relaxation times in {NbRe}-based heterostructures},
  author={F. {De Chiara} and Z. {Makhdoumi Kakhaki} and F. Avitabile and F. Colangelo and A. Kumar and C. Attanasio and C. Cirillo},
  journal={Beilstein J. Nanotechnol.},
  volume={17},
  number={1},
  pages={292},
  year={2026},
  doi={10.3762/bjnano.17.20}
}

@article{battisti2024demonstration,
  title={Demonstration of high-impedance superconducting {NbRe} Dayem bridges},
  author={S. Battisti and J. Koch and A. Paghi and L. Ruf and A. Gulian and S. Teknowijoyo and C. Cirillo and {Z. Makhdoumi Kakhaki} and C. Attanasio and E. Scheer and A. Di Bernardo and G. De Simoni and F. Giazotto},
  journal={Appl. Phys. Lett.},
  volume={124},
  number={17},
  pages={172601},
  year={2024},
  doi={10.1063/5.0200257}
}

@book{tinkham2004introduction,
  title={Introduction to Superconductivity},
  author={Michael Tinkham},
  publisher={Dover Publications},
  address={Mineola, New York},
  year={2004},
  edition={2},
  doi={10.1063/1.2807811}
}

@article{Gri16prb,
  title = {Effective {G}inzburg-{L}andau free energy functional for multi-band isotropic superconductors},
  author = {Konstantin V. Grigorishin},
  journal = {Phys. Lett. A},
  volume = {380},
  number = {20},
  pages = {1781-1787},
  year = {2016},
  issn = {0375-9601},
  doi = {10.1016/j.physleta.2016.03.023},

  langid = {english},
}

@article{Dol08prb,
  title = {Strong electron-phonon interaction in multiband superconductors},
  author = {
    Dolgov, O. V. and 
    Golubov, A. A.},
  journal = {Phys. Rev. B},
  volume = {77},
  issue = {21},
  pages = {214526},
  numpages = {5},
  year = {2008},
  month = {Jun},
  publisher = {American Physical Society},
  doi = {10.1103/PhysRevB.77.214526},

  langid = {english},
}

\end{document}